\pdfoutput=1
\documentclass[journal,10pt]{IEEEtran}  
\IEEEoverridecommandlockouts
\usepackage{cite}
\usepackage{indentfirst}
\usepackage{graphicx}
\usepackage{changepage}
\usepackage{stfloats}
\usepackage{amsfonts,amssymb}
\usepackage{bm}
\usepackage{algorithm}
\usepackage{algorithmic}
\usepackage{amsmath}
\usepackage{amsmath,amssymb,amsfonts}
\usepackage{times}
\usepackage{url}
\usepackage{cite}
\usepackage{textcomp}
\usepackage{xcolor}
\usepackage{color}
\usepackage{setspace}
\usepackage{enumitem}
\usepackage{float}
\usepackage{diagbox}
\usepackage[T1]{fontenc}
\usepackage[utf8]{inputenc}
\usepackage{authblk}
\usepackage{threeparttable}
\usepackage{booktabs}
\usepackage{footnote}
\usepackage{graphicx}
\usepackage{pifont}
\usepackage{subfigure}
\usepackage{mathrsfs}
\usepackage{makecell}
\usepackage{array}
\usepackage{booktabs}
\usepackage{lipsum}
\usepackage{multirow}
\usepackage{adjustbox}
\usepackage{bbding}

\begin{document}
\title{Intelligent Surface Empowered Integrated Sensing and Communication: From Coexistence to Reciprocity}

\author{
	Kaitao Meng, \textit{Member, IEEE}, Qingqing Wu, \textit{Senior Member, IEEE}, Christos Masouros, \textit{Senior Member, IEEE}, Wen Chen, \textit{Senior Member, IEEE}, and Deshi Li
	\thanks{{K. Meng and C. Masouros are the Department of Electronic and Electrical Engineering, University College London, London, UK. }{(email: \{kaitao.meng, c.masouros\}@ucl.ac.uk).}  Q. Wu and W. Chen are with the Department of Electronic Engineering, Shanghai Jiao Tong University, Shanghai 201210, China (emails: \{qingqingwu, wenchen\}@sjtu.edu.cn). D. Li is with the Electronic Information School, Wuhan University, Wuhan, 430072, China (email: dsli@whu.edu.cn). }
}

\maketitle

\begin{abstract}
Integrated sensing and communication (ISAC) has attracted growing interests for sixth-generation (6G) and beyond wireless networks. The primary challenges faced by highly efficient ISAC include limited sensing and communication (S\&C) coverage, constrained integration gain between S\&C under weak channel correlations, and unknown performance boundary. Intelligent reflecting/refracting surfaces (IRSs) can effectively expand S\&C coverage and control the degree of freedom of channels between the transmitters and receivers, thereby realizing increasing integration gains. In this work, we first delve into the fundamental characteristics of IRS-empowered ISAC and innovative IRS-assisted sensing architectures. Then, we discuss various objectives for IRS channel control and deployment optimization in ISAC systems. Furthermore, the interplay between S\&C in different deployment strategies is investigated and some promising directions for IRS enhanced ISAC are outlined.

\end{abstract}


\section{Introduction}
\par
The role of sensing services is poised to become more crucial than ever for the six-generation (6G) wireless networks. To this end, integrated sensing and communication (ISAC) has emerged as a promising candidate technology, in which the unified wireless infrastructures and spectrum resources are utilized for providing both sensing and communication (S\&C) services in an energy, spectrum, and cost-efficient manner \cite{Liu2023Integrated}. ISAC is expected to offer high-throughput and ultra-reliable wireless communication, as well as ultra-accurate and high-resolution wireless sensing \cite{Zhang2021Overview}. Existing research on ISAC predominantly investigates the integration gain to achieve a flexible tradeoff between S\&C performance, such as resource allocation and beampattern design \cite{Cui2021Integrating}. Beyond integration gain, reciprocity between S\&C has great potential to further enhance the performance of both functionalities, producing coordination gains in a mutually beneficial manner \cite{Meng2023Sensing}.

The primary challenges faced by ISAC can be summarized as follows:
1) The S\&C coverage is typically limited due to potential obstructions, particularly for high-frequency signals \cite{Liu2023Integrated};
2) The capacity region of S\&C performance is notably constrained by the correlation between S\&C channels, especially when the channel correlation is weak;
3) The research on coordination gains to improve the performance boundaries of S\&C is still at the preliminary stage.

Recently, intelligent reflecting/refracting surfaces (IRSs), comprising a planar array with numerous passive electromagnetic elements, have been proposed to achieve larger communication coverage and improved transmission quality \cite{Wu2021Tutorial}. In addition to enhancing communication performance, IRSs can increase measurement dimensions and improve sensing accuracy by establishing virtual links and manipulating the channel environment adeptly \cite{fang2023multi}. It is worth noting that the IRS deployment strategies in the literature mainly focus on fixed placement, such as buildings or billboards. Differently, deploying IRSs on mobile equipment, such as vehicles, drones, and even terminals, is another promising solution that can increase the flexibility of S\&C channel control and improve the ability of IRS-mounted targets/devices to be identified/served \cite{Meng2023Sensing}. Moreover, there are several different IRS architectures that can further enhance the performance of S\&C, e.g., incorporating a limited number of receive sensors for channel estimation and environment sensing \cite{Shao2022TargetSensing}. Taking into account these advantages, this work will delve into the exploration of the diverse roles of IRSs within ISAC systems, shedding light on their potential to enhance integration and coordination gains between S\&C.

In this article, we aim to offer a comprehensive overview of the IRS-empowered ISAC systems, identifying key challenges, exploring potential solutions, and pointing out interesting directions to inspire future research. In Section \ref{SensingSufraces}, the fundamental characteristics and innovative IRS-assisted sensing architectures are discussed. Section \ref{SectionJointSandC} presents the new balance consideration of IRS-assisted joint S\&C designs, including fixed-deployed IRS and target-mounted IRS, respectively. Section \ref{SectionSensingAssistedC} explores the reciprocity between S\&C with various sources of performance enhancement. Section \ref{Extensions} outlines promising future directions in ISAC enhanced by IRS. Finally, \ref{Conclusions} concludes this work.

\section{Intelligent Surfaces Enhanced Sensing}
\label{SensingSufraces}
While extensive research has explored different applications of IRSs across diverse wireless communication scenarios, the operational principles of IRS-assisted sensing exhibit considerable variation, leaving many open issues worthy of study. In the subsequent discussion, we will investigate the fundamental channel characteristics of IRS-aided sensing, presenting several efficient architectures based on these insights. 

\begin{figure*}[t]
	\centering
	\includegraphics[width=13cm]{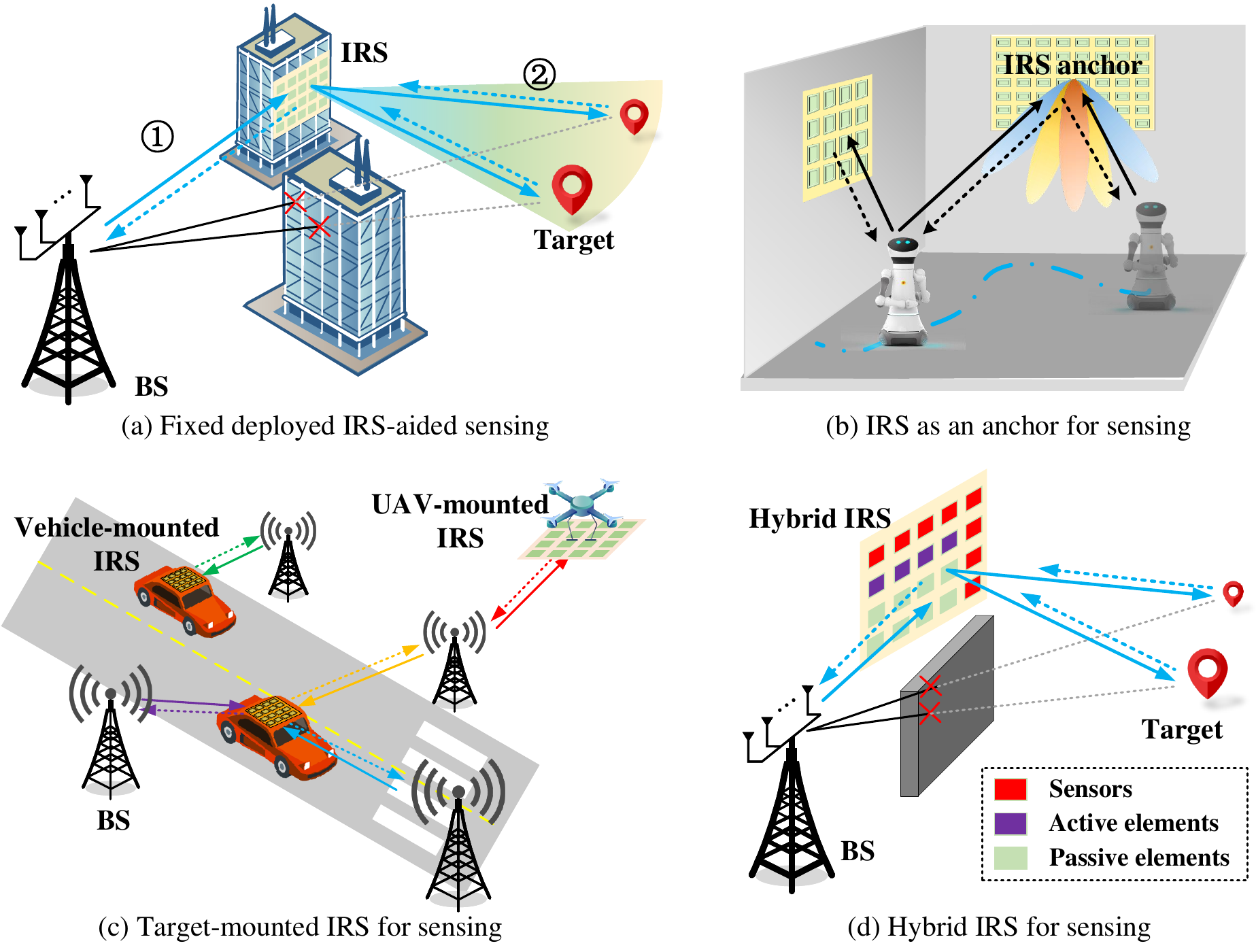}
	\caption{Illustration of new sensing framework aided by IRSs.}
	\label{figure2}
\end{figure*}

\subsection{Fundamental Characteristics of IRS-aided Sensing} 
\label{SensingFundamental}
One of the key distinctions of passive IRS-aided sensing is its unique role in obtaining the channel parameters of the target. In particular, aside from enhancing the power of the echo signals reflected from a target, there are specific requirements regarding the measurement direction and the degree of freedom (DoF) of channels between the target and the sensing receiver (e.g., base stations (BSs) and radar). 

\subsubsection{Channel Requirements} 
\label{ChannelRequirements}
As shown in Fig.~\ref{figure2}(a), the IRS can establish a virtual Line-of-Sight (LoS) link to realize target detection and parameter estimation in blind areas by utilizing prior information on transceiver-IRS channels \cite{Meng2022Intelligent, Wu2021Tutorial}. 
Nevertheless, unlike IRS-assisted communication systems, achieving effective target parameter estimation, even for single-target cases, typically necessitates a high-rank channel between the transceiver and the IRS, i.e., part $\textcircled{1}$ in Fig.~\ref{figure2}(a). This requirement arises from the challenge of accurately conveying multiple target parameters (e.g., reflection coefficient and angle information) through a rank-one channel \cite{Song2023Intelligent}, in which case the estimation performance severely degrades. By contrast, a LoS channel is required between the IRS and the target, i.e., part $\textcircled{2}$ in Fig.~\ref{figure2}(b), with non-LoS (NLoS) links often deemed as undesirable interference. Therefore, how to effectively design the phase shift and IRS placement to achieve optimal sensing performance still needs further investigation.

\subsubsection{Shared Links of Multiple Targets} 
In IRS-aided sensing scenarios with multiple targets, the IRS-transceiver channel accommodates echo signals reflected from multiple targets, leading to identical echoes' directions observed at the sensing receiver, as illustrated in Fig.~\ref{figure2}(a). Therefore, in practice, it is challenging to effectively sense multiple targets simultaneously via the same virtual LoS link. To tackle this issue, a novel coding strategy was proposed in \cite{Meng2022Intelligent}, where a correlation between the target directions and the signal sequence is established to distinguish targets with the help of echo signal analysis. By adopting such an encoding method, unlike traditional radars that mainly perform data processing at the receiver side, the proposed scheme in \cite{Meng2022Intelligent} actively adopts a signal pre-processing strategy at the transmitter side to assist in distinguishing targets and speeding up the data processing efficiency at the receiver side. While employing multiple IRSs to achieve highly efficient multi-target sensing can enhance coverage and measurement accuracy, it is worth noting that identifying and distinguishing signals reflected from different targets will become more challenging, which deserves further investigation.

\subsection{New Architectures}
In addition to the IRS coverage enhancement methods discussed above, there are some new architectures for more advanced IRS-aided sensing, as shown in Figs.~\ref{figure2}(b)-\ref{figure2}(d). 

{\bf{IRS anchor}}: IRS can work as a standalone anchor to provide localization services without relying on other anchors, as illustrated in Fig. \ref{figure2}(b). Specifically, when a nearby target requires positioning services, it can direct beams towards the IRS, and then the IRS dynamically adjusts the phase shifts to change the echo's direction. In this case, the target's direction relative to the IRS can be estimated based on the power of the echo signals received at the target. A direct method to design IRS phase shift is utilizing the scanning strategy, achieved by changing the IRS phase shifts in various possible directions. Then, the direction index with the largest received power corresponds to the target direction. Nonetheless, such a self-anchoring system may pose new challenges to resolve, e.g., the complicated balance between localization accuracy and scanning overhead. To handle this issue, a possible solution is the application of the hierarchical codebook-based IRS reflection design to realize a flexible tradeoff between accuracy and efficiency.

{\bf{Target-mounted IRS}}: IRS-mounted targets can actively focus the echoes in one or more desired directions according to the location of the sensing transceivers, as shown in Fig. \ref{figure2}(c). This effect is particularly pronounced for targets with a relatively small radar cross section (RCS) \cite{Shao2022TargetSensing}. Furthermore, in the case with extended targets, their positions and orientations can be accurately calculated by strategically placing multiple IRSs on the surfaces of these targets. However, owing to the inherent difficulty of accurately determining the IRS's position, designing the IRS phase shift to reliably direct the echo signal toward the intended sensing receiver remains a formidable challenge, which deserves further study. In addition to enhancing localization performance, the phase shift control of IRSs can also help to modulate information on the echo signals to achieve identifier correlation of multiple targets, which is practically useful for target association in dense target scenarios.

{\bf{Hybrid IRS}}: Incorporating the active elements with power amplifying function and sensors at the IRS in a hybrid setup enhances its ability for sensing coverage and echo signal analysis in the blind area, as illustrated in Fig. \ref{figure2}(d). In this case, beyond mere data processing on the BS side, there are two links that can be used for the target's information analysis: the BS-IRS-target-sensors link and the BS-IRS-target-IRS-BS link. While integrating sensors into the IRS effectively reduces path loss of echo signals, it is practically difficult to meet the high timeliness requirements of sensing results due to the IRS's limited computing ability. A viable solution is to transmit the results back to the BS after performing basic preprocessing on the IRS side. Moreover, the data fusion of these two sensing links raises critical concerns, including determining optimal data processing strategy and striking an effective balance in the deployment of passive/sensors/active elements. 

For the above three sensing frameworks, deploying multiple IRSs can further create additional LoS opportunities with targets, but the challenge arises when attempting to gather global information from all IRSs for cooperative localization due to the excessive overhead and complexity. One promising solution to address this concern is the development of efficient distributed sensing methods, with the sharing of low-dimensional local information among themselves. However, this approach may incur a noticeable performance loss due to incomplete data collection. Particularly, how to realize efficient and reliable sensing data exchange and fusion among multiple IRSs for high-quality and seamless-coverage sensing is an open problem, deserving further investigation.

\begin{table*}[t] 
	\centering
	\caption{Illustration of IRS roles in S\&C channel control for various scenarios ({\ding{52}}: Yes; {\ding{55}}: No; $\star$: depends on tasks). }
	\label{Table1}
	\begin{tabular}{ |p l | l | l| l | l| l|}
			\hline
			\multicolumn{2}{|c|}{{  \diagbox[]{Applications}{Objectives}   }} & {\makecell[c]{Increasing\\
			channel gain}} & {\makecell[c]{Increasing\\ correlation}} & {\makecell[c]{Decreasing\\ correlation}} & {\makecell[c]{High\\ rank}}  \\ \cline{2-6}
			\hline
			\multicolumn{1}{|c|}{\multirow{3}{*}{\begin{tabular}[c]{c}Sensing\end{tabular}}} & {\makecell[c]{Transmitter-IRS-target}}  & {\makecell[c]{\ding{52}}} & {\makecell[c]{\ding{52} }} & {\makecell[c]{{\ding{55}} }} & {\makecell[c]{{\ding{55}}}} \\ \cline{2-6}
			\multicolumn{1}{|c|}{}   & {\makecell[c]{Target-IRS-receiver}}  & {\makecell[c]{ \ding{52} }} & {\makecell[c]{{\ding{55}} }}  & {\makecell[c]{ \ding{52} }}  & {\makecell[c]{ \ding{52}\\ }} \\ 
			\hline
			\multicolumn{1}{|c|}{\multirow{2}{*}{\begin{tabular}[c]{@{}c@{}}Communication\end{tabular}}} & {\makecell[c]{Multicast}}  & {\makecell[c]{\ding{52} }} & {\makecell[c]{\ding{52} }} & {\makecell[c]{{\ding{55}} }}  & {\makecell[c]{{\ding{55}} }} \\ \cline{2-6}
			\multicolumn{1}{|c|}{}   & {\makecell[c]{Unicast}}  & {\makecell[c]{\ding{52}}} & {\makecell[c]{\ding{55}}} & {\makecell[c]{\ding{52}}} & {\makecell[c]{\ding{52}}} \\ 
			\hline
			\multicolumn{1}{|c|}{\multirow{2}{*}{\begin{tabular}[c]{@{}c@{}}ISAC\end{tabular}}} & {\makecell[c]{Single user/target}}  & {\makecell[c]{\ding{52} }} & {\makecell[c]{\ding{52} }} & {\makecell[c]{{\ding{55}}}}  & {\makecell[c]{$\star$ }} \\ \cline{2-6}
			\multicolumn{1}{|c|}{}   & {\makecell[c]{Multiple users/targets}}  & {\makecell[c]{\ding{52}}} & {\makecell[c]{\ding{52}}} & {\makecell[c]{$\star$}} & {\makecell[c]{\ding{52}}} \\ 
			\hline
	\end{tabular}
\end{table*} 

\section{Balance For IRS-empowered ISAC}
\label{SectionJointSandC}

\subsection{Channel Balance for Sensing and Communication}
The design of IRS phase shifts needs to be tailored to achieve distinct objectives in channel control across various applications, including communication-only, sensing-only, and S\&C scenarios, as illustrated in Table \ref{Table1}, thereby bringing a new balance in IRS-assisted ISAC systems.

\subsubsection{IRS-aided Communication Channel}

When the BS broadcasts the same data to multiple users, the IRS mainly focus on introducing supplementary channel paths to amplify the effective channel gains from the BS to all these users. Differently, when the BS allocates the same time-frequency resource block to send independent data signals to multiple users, besides enhancing effective channel gain, the IRS phase shift design should aim to reduce channel correlation between different users, thereby suppressing multi-user interference. The main reason is that channels highly correlated between users will lead to limited multiplexing gain in multi-user communication, particularly within the high signal-to-interference-plus-noise ratio (SINR) region. For instance, when the IRS is able to manipulate inter-user channels to achieve orthogonality \cite{Alegra2022Orthogonalization}, the optimal BS transmit beamforming solution becomes maximum-ratio transmission (MRT), leading to a reduced complexity in precoding design. In addition to the above-mentioned role of IRS in single cells, it can also be used to suppress inter-cell interference in multi-cell wireless networks \cite{Khan2023IntegrationNOMA}.

\subsubsection{IRS-aided Sensing Channel}

In the case of the IRS-assisted sensing model, the IRS's role in channel control can be primarily categorized into two types based on its contribution to the sensing link: when the IRS is positioned between the sensing transmitter and the target, aimed at enhancing the target's illumination; the IRS is deployed between the sensing receiver and the target, focused on facilitating the analysis of target echoes \cite{Foundations2022Buzzi}.
In particular, for parameter estimation (e.g., angle and RCS), it is essential to ensure sufficient DoF in the channel between the receiver and the IRS, guaranteeing that the channel can effectively convey all the required measurement parameters \cite{Song2023Intelligent}. Moreover, for multi-target sensing, an extra aim of IRS phase shift design is to minimize the correlation between distinct target channels, enabling effective differentiation of echoes reflected from different targets, as summarized in Table \ref{Table1}. For example, in \cite{Meng2022Intelligent}, to effectively distinguish multiple targets' echoes via the shared IRS-BS links, the correlation between the channels of different targets is reduced to facilitate the simultaneous detection of multiple targets. 

\subsubsection{IRS-aided ISAC Channel}
The ISAC system can achieve more enhanced performance with stronger coupling between S\&C channels, allowing for more efficient utilization of unified signals for S\&C tasks. Therefore, through optimizing phase adjustments of the IRS, a weakly coupled S\&C subspace can be rotated to manifest coupling, bringing substantial enhancements in the tradeoff between S\&C performance when IRSs are integrated. However, considering the weak channel correlation requirements for multi-user communications and the high-rank channel demand for target parameter estimation, a key challenge for IRS-assisted ISAC scenarios is how to simultaneously achieve these different control objectives for S\&C tasks. Moreover, the difference in channel coherence time between users and targets gives rise to varying phase shift control frequency demands for S\&C, necessitating a more efficient phase shift design strategy that capitalizes on this frequency disparity.

In addition, for target-mounted IRSs, it is possible to achieve a new vehicle localization and communications tradeoff by balancing the power/time resources of the received signal at IRSs for refraction and reflection \cite{Meng2023Sensing}. Specifically, when the IRS's phase shift design prioritizes maximizing the power over a longer period at the sensing receiver, it enhances target localization accuracy. By contrast, when more signal energy is refracted towards the devices within the vehicle, it brings improved communication performance.

\subsection{Deployment Balance Between S\&C}
In this subsection, we mainly discuss two IRS deployment strategies and their corresponding balance between S\&C: fixed deployment and target mounting. 

\subsubsection{Fixed Deployment Intelligent Surfaces}
\begin{figure*}[t]
	\centering
	\includegraphics[width=14.4cm]{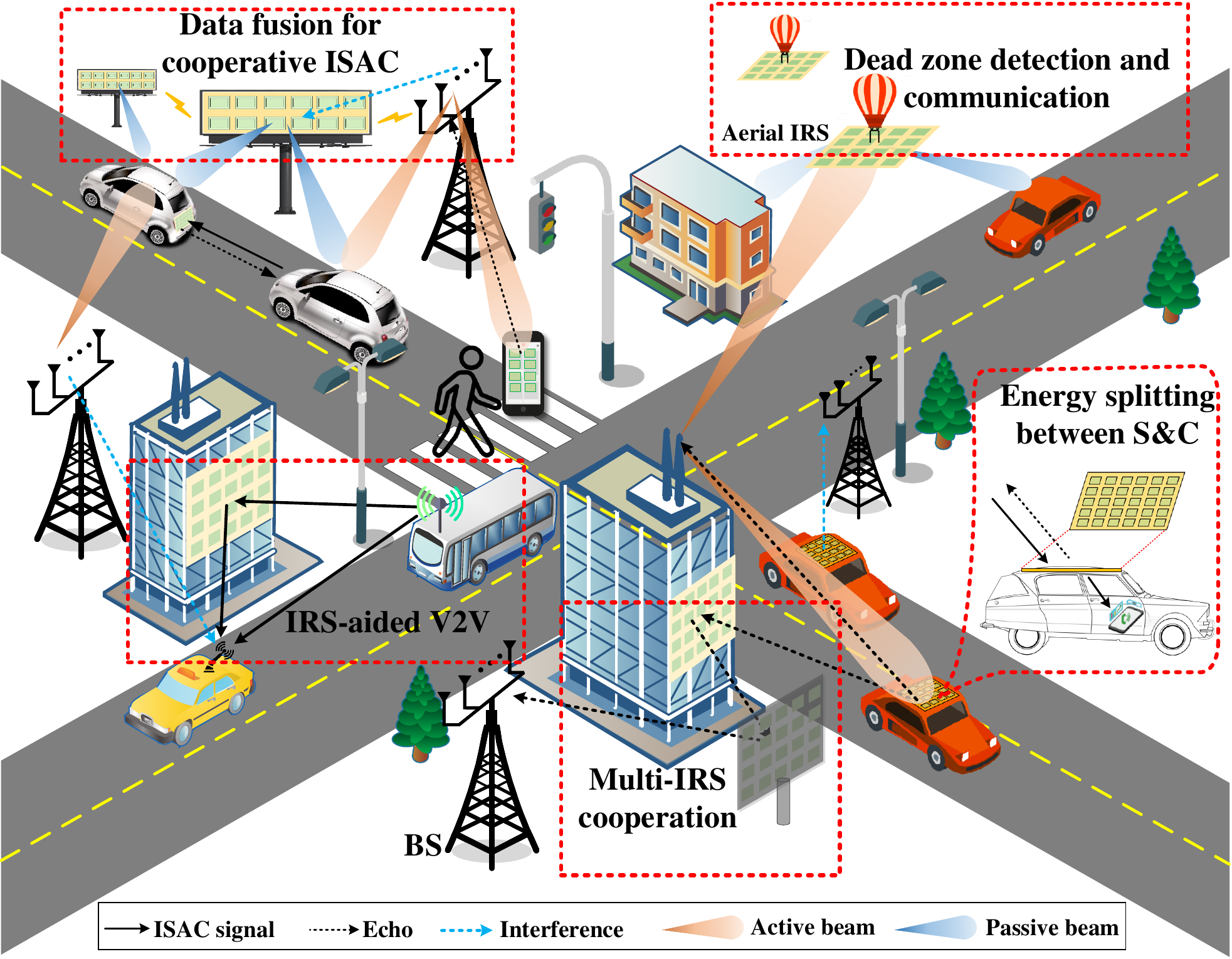}
	\caption{Scenarios of ISAC networks empowered with multiple IRSs.}
	\label{figure1}
\end{figure*}

The optimal fixed deployment strategy of IRSs in wireless networks to maximize the network capacity is discussed in \cite{Wu2021Tutorial}. However, unlike communication, sensing operations primarily rely on (IRS-established virtual) LoS links between targets and sensing receivers. Accordingly, an IRS positioned at a higher altitude and open environment is more inclined to establish strong LoS links with targets, thereby facilitating the utilization of a larger quantity of reflected signals by nearby sensing receivers. Conversely, the sensing quality may suffer from the poor DoF of the LoS-dominated links between IRSs and sensing receivers, as discussed in Section \ref{ChannelRequirements}. Hence, the ideal deployment location for IRS-assisted sensing necessitates a robust LoS connection with the target and a high-rank multiple-input and multiple-output (MIMO) channel with the sensing receiver. Furthermore, deploying multiple IRSs in ISAC networks is a promising solution to achieve larger S\&C coverage and significantly improve both S\&C performance, as shown in Fig.~\ref{figure1}, which is a new problem of high practical interest.

\subsubsection{Target-mounted Intelligent Surfaces}
From the perspective of target sensing, it is demonstrated that the RCS of the IRS can be effectively changed by controlling its phase shifts \cite{Shao2023National}. Therefore, deploying IRS on specific targets can either boost echo power at the legitimate receiver to enhance its localization accuracy or attenuate echo power at the unauthorized station to guarantee sensing security. In addition, targets such as vehicles are generally deemed as extended targets, and in this case, deploying multiple IRSs at distinct and predefined locations, such as the four corners of the vehicle, can significantly enhance the acquisition of its attitude information, such as azimuth angle, rotation speed, etc. In contrast, centralized IRS can maximize passive beamforming gain, which can enhance echo power received at the IRS and further improve the communication performance of users inside the vehicle. Consequently, a fundamental tradeoff between target sensing accuracy and vehicular user communication performance emerges, rendering it an ongoing area of research.


\section{Reciprocity Between Sensing and Communication}
\label{SectionSensingAssistedC}

In this section, we provide three paradigms of reciprocity between S\&C with the assistance of IRSs in ISAC systems, including sensing-assisted communication, communication-assisted sensing, and reciprocity between S\&C, and discuss the corresponding IRS roles, advantages, and challenges, as summarized in Table \ref{Table2}. 

\subsection{Sensing-assisted Communication}
\label{SensingAssistedCommunication}
Based on the source of communication performance gain brought by sensing, we mainly discuss two categories of sensing-assisted communication schemes empowered by IRSs: those leveraging sensing results for communication (e.g., angle and range) and those utilizing sensing resources for communication (e.g., sensing waveforms).  

\begin{table*}[t] 
	\centering
	\caption{IRS roles for Reciprocity between S\&C.}
	\label{Table2}
	\begin{tabular}{ |p l | l | l| l | l|}
		\hline
		\multicolumn{1}{|c|}{{  {Paradigms} }} & {\makecell[c]{Source of\\coordination gain}}  & {\makecell[c]{IRS roles}} & {\makecell[c]{Advantages}} & {\makecell[c]{Challenges}}  \\ \cline{1-5}
		\hline
		\multicolumn{1}{|c|}{\multirow{2}{*}{\begin{tabular}[c]{c}Sensing-assisted\\communication\end{tabular}}} & {\makecell[c]{Sensing results}}  & {\makecell[c]{Improve echo power,\\channel estimation\\for semi-passive IRSs}} & {\makecell[c]{Overhead reduction,\\reliable communication,\\enhance echo power}} & {\makecell[c]{Target association, high\\timeliness requirements\\}} \\ \cline{2-5}
		\multicolumn{1}{|c|}{}   & {\makecell[c]{Sensing resource}}  & {\makecell[c]{Improve channel quality,\\modulation data at IRSs\\}}  & {\makecell[c]{ Energy efficiency\\improvement\\ }} & {\makecell[c]{Discontinuous resource }} \\ 
		\hline
		\multicolumn{1}{|c|}{\multirow{2}{*}{\begin{tabular}[c]{@{}c@{}}Communication-\\assisted sensing\end{tabular}}} & {\makecell[c]{Communication\\data}} & {\makecell[c]{Assist data transmission\\ and data fusion }} & {\makecell[c]{Improve sensing\\efficiency}}  & {\makecell[c]{ Time synchronization}} \\ \cline{2-5}
		\multicolumn{1}{|c|}{}   & {\makecell[c]{Communication\\resource}}  & {\makecell[c]{Estimate signal direction,\\extend sensing coverage}} & {\makecell[c]{Increased sensing\\diversity}}  & {\makecell[c]{Limited signal\\processing ability}} \\ 
		\hline
		\multicolumn{2}{|c|}{{Reciprocity between S\&C}} & {\makecell[c]{Balance between S\&C,\\control channel correlation,\\enhance coordination gain}}  & {\makecell[c]{More flexible\\tradeoff, higher\\integration gain}} & {\makecell[c]{Describe accurate\\performance boundary}}  \\ 
		\hline
	\end{tabular}
\end{table*}

\subsubsection{Sensing Results for Communication}

The sensing results (user location and environmental information) can be exploited to facilitate the IRS phase shift design and BS resource allocation, thereby improving communication performance. For example, an IRS passive beam tracking mechanism can be implemented by analyzing the users' location information based on the echoes reflected from users, all without incurring any signalling overhead. It is also possible to predict potentially blocked areas with the help of environmental information, enabling IRSs to establish supplementary channels in advance. This greatly reduces the outage probability, especially for high-mobility communication users, thereby achieving ultra-reliable data transmission. Additionally, semi-passive IRS (refers to the IRS incorporated with sensors) can leverage its sensory data to predict the positions of nearby users requiring services, thus assisting in the IRS phase shift configurations and BS beamforming optimization. As a result, signalling overhead is reduced significantly and a low-latency communication service is more likely to be provided.

For IRS-mounted vehicles, sensing results can also be leveraged to enhance the communication performance for users within the vehicle. For instance, in \cite{Meng2023Sensing}, a novel two-stage ISAC protocol for vehicular communication networks was proposed. This protocol includes a joint S\&C stage and a communication-only stage, where the state estimation and measurement results from the joint S\&C stage are efficiently exploited to optimize IRS phase shifts for communication enhancement in the subsequent communication-only stage. In practice, reflections from various vehicles can lead to mutual interference among different sensing transceivers, as shown in Fig.~\ref{figure1}. Therefore, the network positioning performance critically relies on the effectiveness of utilizing the beam steering capabilities of IRSs to actively establish the correlation between targets and transceivers, thereby facilitating target association. 

\subsubsection{Sensing Resource for Communication}
Utilizing time/frequency/space division multiple access techniques, communication data can be modulated on the sensing beam to send messages to the receiver \cite{Liu2022Integrated}. In this case, the IRS can not only improve the effective channel gain from the transmitter to the receiver but can also modulate sensory data on the sensing signals to enable low-power data transmission of the IRS. Considering that modulating data on the IRS impacts the communication rate at the transmitter side and elevates the complexity of data demodulation at the receiver, a delicate balance is required in managing the phase control of the IRS to optimize both the transmitter's and IRS's data transmission performances. Moreover, previous ISAC studies primarily concentrate on assessing the S\&C performance throughout the entire considered period. However, sensing tasks are generally performed at regular intervals, such as periodic target detection in areas of interest, routine updates of the environment status, etc. These periodic sensing resources are well-suited for providing delay-tolerant communication services, such as Internet of Things (IoT) data collection. However, utilizing such regular sensing resources to ensure the quality of time-sensitive communication tasks presents new challenges and is worthy of further research.

\subsection{Communication-assisted Sensing}
In this subsection, based on the source of the sensing performance gain, two kinds of communication-assisted sensing schemes are discussed: communication data for sensing, and communication waveform for sensing.

\subsubsection{Transmission Data for Sensing}
Sensing performance is determined not only by the quality of the original sensory data, but also by data transmission performance in practice, such as transmission delay and distortion rate, especially for remote sensing scenarios. In this case, the IRS-empowered sensing system brings a fundamental tradeoff in sensing data acquisition and transmission under limited spectral resources, which has not been studied yet and deserves further investigation. Moreover, the data fusion performance in multiple sensor systems also depends heavily on effective wireless communication techniques, e.g., the accuracy of multi-static sensing systems. An appealing strategy for enhancing data fusion efficiency for the above-mentioned sensing scenarios is to utilize over-the-air computation \cite{Fang2021OvertheAir}. This method capitalizes on the inherent waveform superposition capability of wireless channels, enabling the simultaneous transmission of aggregated data by multiple sensors, and eliminating the need for separate data demodulation and fusion processes. In this case, IRSs can adjust the phase shifts according to the sensor channel to improve the quality and timeliness of sensory data transmission. However, for distributed large-scale sensing systems, the deployment and phase shift optimization of IRSs poses significant challenges due to the non-negligible transmission delay differences of multiple sensors and the complicated interference between various sensing systems, which deserves further study.

\subsubsection{Communication Resource for Sensing}
Various communication signals from BSs and users can be used for sensing tasks, including pilot signals and data signals. Pilot signals transmitted by BSs generally employ an omnidirectional waveform to facilitate wide-ranging channel acquisition, whereas the waveforms of data signals are customized to match the unique characteristics of users' channels. Therefore, the challenge lies in how the IRS can continuously harness the dynamic waveforms of these signal sources to maintain stable performance in sensing. To overcome this issue, one potential solution is to employ the semi-passive IRS for statistical analysis of data source directions and then fine-tune IRS phase shifts to steer the beam toward the intended sensing area or target. Moreover, for IRS-mounted targets, the IRS can reflect the communication signal in the direction of the sensing receiver to facilitate their localization. In practice, BSs may be deployed in a distributed manner, leading to dispersed communication data sources and sensing receivers. In this case, how to design the IRS phase shift for precise control of echo signals towards several distributed sensing receivers is a captivating and unresolved challenge.

\subsection{Mutual Assistance Between Sensing and Communication}
When the above two reciprocity between S\&C mechanisms are employed at the same time, S\&C tasks can be innovatively designed to assist each other for achieving a more efficient win-win integration. Considering that performance improvements in S\&C appear to be intertwined with each other, one challenging issue is theoretically characterizing the boundaries of S\&C performance with mutual assistance. One possible solution is to analyze the outer performance bound. For instance, without taking mutual assistance into account, we denote the S\&C performance as $S_1$ and $C_1$, respectively, when all resources are dedicated solely to sensing or communication. An outer bound can be established through the following process. Leveraging the sensing results from $S_1$ to enhance communication leads to the upper bound of communication performance, denoted as $C_2$. Likewise, utilizing the communication outcomes from $C_1$ to improve sensing yields the upper bound of sensing performance, represented by $S_2$. In practice, due to the coupling of S\&C resources, it is difficult to achieve optimal S\&C performance $S_2$ and $C_2 $ even with mutual assistance between S\&C. 

Another interesting question is whether sensing or communication always leads to performance improvements for each other. Note that if we only focus on optimizing the performance of communication (sensing), incorporating additional sensing (communication) tasks may not necessarily enhance the system performance, as it leads to extra resource consumption. For instance, \cite{Meng2023Sensing} provided a sufficient and necessary condition on whether the sensing is needed or not in IRS-assisted vehicle ISAC systems to improve the data transmission performance. Specifically, when the accuracy of estimated angles is higher or the accuracy of measured angles is lower, i.e., the communication performance gain brought by sensing cannot compensate for the performance loss caused by the time and power consumption. Under these circumstances, opting for pure communication (sensing) is the most favourable choice. Furthermore, how to explore the effective utilization of IRS for interference suppression and collaborative design of S\&C tasks in network-level ISAC is a promising issue for future research. In addition, the exploration of leveraging IRSs to enhance interference suppression and facilitate collaborative S\&C tasks within network-level ISAC is a question worthy of further study.

\subsection{Case Study: Mutual Assistance Between Sensing and Communication}
\begin{figure}[t]
	\centering
	\includegraphics[width=8.4cm]{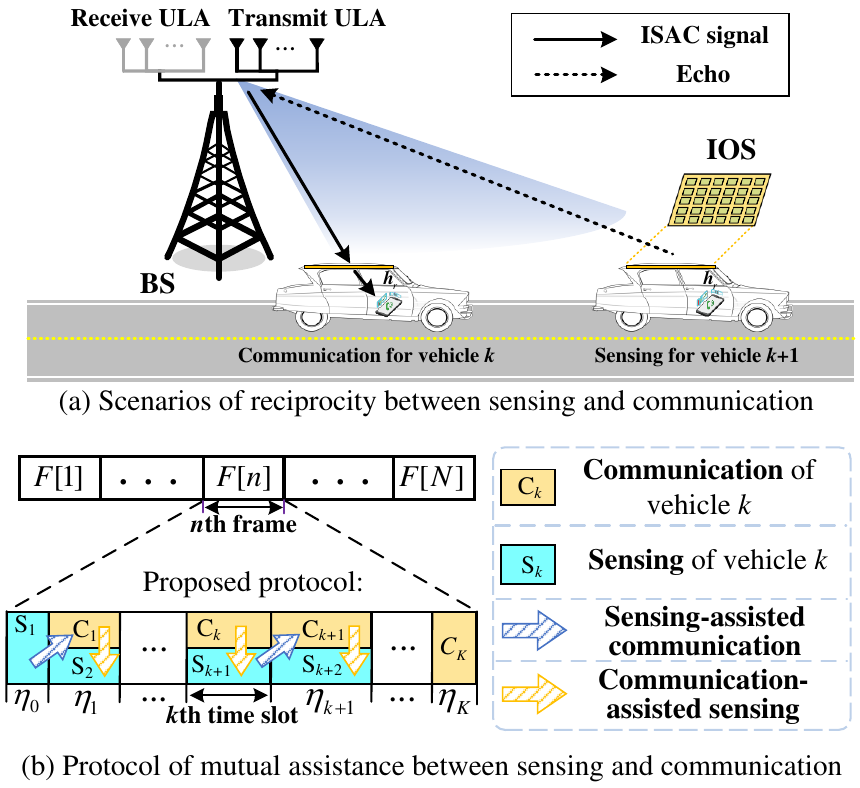}
	\caption{Illustration of IRS-empowered ISAC systems.}
	\label{figure4}
\end{figure}

As illustrated in Fig.~\ref{figure4}, we explore an IRS-aided ISAC system for mutual assistance between S\&C. In this setup, one BS transmits independent signals to several mobile vehicles equipped with IRS, employing a time-division approach and analyzing the echo signals simultaneously reflected by the IRS for localization. The BS employs a single transmit antenna and a general uniform linear array (ULA). The vehicles are assumed to travel along a straight road, with each vehicle equipped with an IRS featuring both reflection and refraction modes. As illustrated in Fig.~\ref{figure4}, an example of the synergy between S\&C is provided as follows.
\begin{itemize}
	\item {\textit{Sensing-assisted communication}}: During the $k$th time slot, the BS transmits signal $s_k$ to vehicle $k$, while IRS $k$ operates in the \textit{refracting} mode, with its phase shifts designed for communication according the sensing results of time slot $k-1$.
	\item {\textit{Communication-assisted sensing}}: Signal $s_k$ arriving at IRS $k+1$ during the $k$th time slot is deemed redundant; however, it can be harnessed to amplify echo signals and assist in localizing vehicle $k+1$ by configuring IRS $k+1$ into the \textit{reflecting} mode.
\end{itemize} 
 
\begin{figure}[t]
	\centering
	\includegraphics[width=7cm]{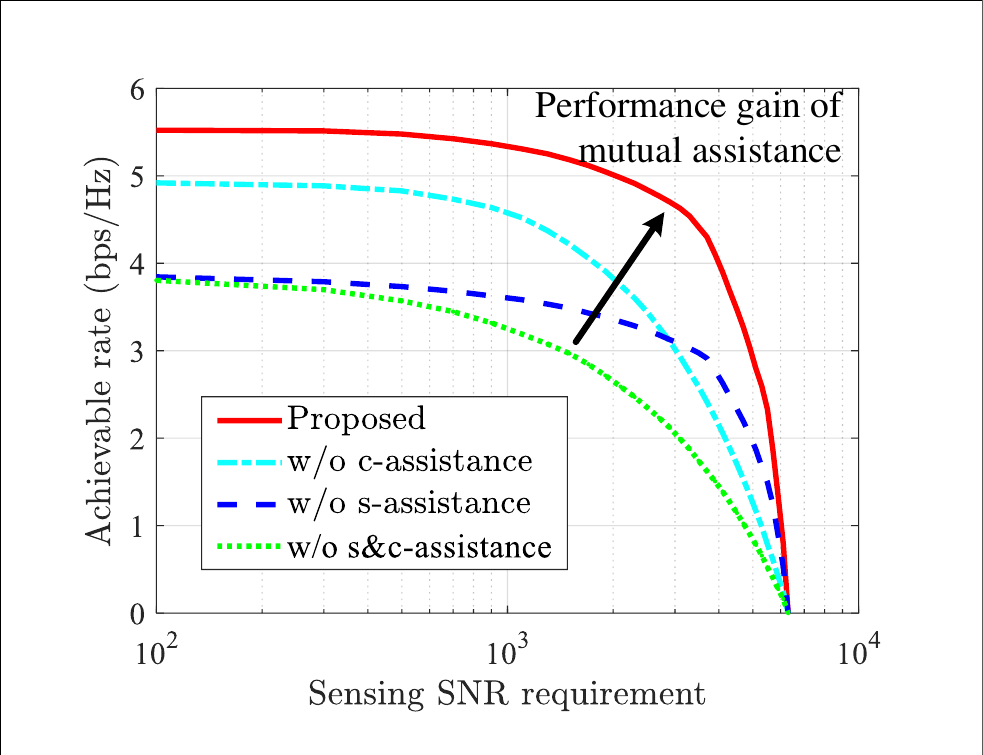}
	\caption{Performance enhancement brought by mutual assistance between S\&C.}
	\label{figure3}
\end{figure}

Based on the above design, the echo signal generated by interference provides new opportunities to improve localization performance. To verify the effectiveness of reciprocity between S\&C, the proposed scheme is compared with three benchmarks: Without sensing assistance (w/o s-assistance); without communication assistance (w/o c-assistance); without S\&C assistance (w/o s\&c-assistance). In Fig.~\ref{figure3}, it can be observed that the communication rate, as obtained by all schemes, decreases as the sensing SNR requirement increases. This observation aligns with expectations, as larger SNR requirement values impose more stringent constraints on the resource allocated for communication. 
From Fig.~\ref{figure3}, it is evident that the achievable rate of the proposed scheme surpasses that of the scheme without sensing assistance, particularly when the sensing requirement is lower. This is attributed to the optimized time allocation in the proposed scheme, thereby amplifying the communication benefit derived from sensing. Furthermore, at a fixed achievable rate of 4.9 bps/Hz, the sensing performance of the proposed scheme markedly outperforms that of the scheme without communication assistance. This is primarily due to the increased utilization of information signals, which aids in localization and consequently reduces the time required for dedicated sensing.

\section{Other Extensions}
\label{Extensions}
Some open issues and challenges related to ISAC with IRSs are discussed as follows.

\subsection{Cooperation between Multiple IRSs}  

Multiple distributed IRSs can be strategically deployed to assist BSs in offering ISAC services within distinct LoS-blocked regions. Utilizing the same total number of IRS elements, in comparison to centrally deployed IRS, while distributed IRSs naturally decrease the power of echoed signals due to a reduced aperture, it can largely expand the coverage area and enhance sensing diversity. Consequently, there arises a tradeoff concerning the division sizes of distributed IRSs, which deserves further study.

\subsection{Near-field ISAC with IRSs}
Compared to the conventional far-field system, near-field ISAC introduces an additional distance dimension to allocate resources and improve performance for both S\&C. In turn, the estimation of user channel and target parameters becomes more complicated due to the new involved distance domain. In this case, with the assistance of IRSs in ISAC scenarios, it motivates new techniques to design BS transmit beamforming and IRS phase shift for achieving better integration/coordination gains between S\&C.

\subsection{Artificial Intelligence for IRS-assisted ISAC}
In the context of distributed machine learning architecture, the incorporation of IRSs can enhance the efficiency of parameter learning and accelerate algorithm convergence. Furthermore, IRSs play a crucial role in enhancing the accuracy and quality of sensing data, leading to a substantial improvement in machine learning efficiency. Consequently, within the unified ISAC learning framework, IRSs need to strike a balance in terms of learning gradient transmission and sensing data acquisition.

\subsection{Secure ISAC assisted by IRSs}
Security is another critical challenge for ISAC networks due to the inherent broadcast nature of wireless signals. The existing IRS secure research primarily focuses on exploiting IRSs to prevent eavesdroppers from acquiring information. Moreover, mounting the IRS on the target allows precise control of the reflected echo signals. Thus, the IRS phase shift design enables the management of received power for the intended sensing receivers while minimizing the power at potential eavesdroppers. While IRSs provide a more flexible balance in simultaneously achieving secure S\&C, the high complexity of obtaining the locations and channels of potential eavesdroppers makes securing ISAC services particularly challenging.

\section{Conclusions}
\label{Conclusions}

In this article, we have delved into the potential of IRSs to enhance ISAC systems, aiming to unveil the integration benefits and reciprocal relationships between S\&C. We shed light on novel design considerations and highlighted essential challenges when incorporating IRS in ISAC setups. Particularly, we emphasize practical applications demonstrating the coordination gain in ISAC, such as sensing-assisted communication, communication-assisted sensing, and mutual assistance between S\&C, to demonstrate that S\&C can be implemented in a complementary way. Furthermore, we validate these concepts through a complete case study. Given the relatively uncharted territory of IRS-empowered ISAC, this paper aims to serve as a useful reference for future research in this domain.

\footnotesize  	
\bibliography{mybibfile}

\begin{thebibliography}{10}
\providecommand{\url}[1]{#1}
\csname url@samestyle\endcsname
\providecommand{\newblock}{\relax}
\providecommand{\bibinfo}[2]{#2}
\providecommand{\BIBentrySTDinterwordspacing}{\spaceskip=0pt\relax}
\providecommand{\BIBentryALTinterwordstretchfactor}{4}
\providecommand{\BIBentryALTinterwordspacing}{\spaceskip=\fontdimen2\font plus
\BIBentryALTinterwordstretchfactor\fontdimen3\font minus
  \fontdimen4\font\relax}
\providecommand{\BIBforeignlanguage}[2]{{%
\expandafter\ifx\csname l@#1\endcsname\relax
\typeout{** WARNING: IEEEtran.bst: No hyphenation pattern has been}%
\typeout{** loaded for the language `#1'. Using the pattern for}%
\typeout{** the default language instead.}%
\else
\language=\csname l@#1\endcsname
\fi
#2}}
\providecommand{\BIBdecl}{\relax}
\BIBdecl

\bibitem{Liu2023Integrated}
R.~Liu, M.~Li, H.~Luo, Q.~Liu, and A.~L. Swindlehurst, ``Integrated sensing and
  communication with reconfigurable intelligent surfaces: Opportunities,
  applications, and future directions,'' \emph{{IEEE} Wireless Commun.},
  vol.~30, no.~1, pp. 50--57, Feb. 2023.

\bibitem{Zhang2021Overview}
J.~A. Zhang, F.~Liu, C.~Masouros, R.~W. Heath, Z.~Feng, L.~Zheng, and
  A.~Petropulu, ``An overview of signal processing techniques for joint
  communication and radar sensing,'' \emph{IEEE J. Sel. Top. Signal Process.},
  vol.~15, no.~6, pp. 1295--1315, Nov. 2021.

\bibitem{Cui2021Integrating}
Y.~Cui, F.~Liu, X.~Jing, and J.~Mu, ``Integrating sensing and communications
  for ubiquitous {IoT}: Applications, trends, and challenges,'' \emph{IEEE
  Netw.}, vol.~35, no.~5, pp. 158--167, Sep./Oct. 2021.

\bibitem{Meng2023Sensing}
K.~Meng, Q.~Wu, W.~Chen, and D.~Li, ``Sensing-assisted communication in
  vehicular networks with intelligent surface,'' \emph{{IEEE} Trans. Veh.
  Commun.}, pp. 1--17, 2023.

\bibitem{Wu2021Tutorial}
Q.~Wu, S.~Zhang, B.~Zheng, C.~You, and R.~Zhang, ``Intelligent reflecting
  surface-aided wireless communications: A tutorial,'' \emph{{IEEE} Trans.
  Commun.}, vol.~69, no.~5, pp. 3313--3351, May 2021.

\bibitem{fang2023multi}
Y.~Fang, S.~Zhang, X.~Li, J.~Xu, and S.~Cui, ``Multi-{IRS}-enabled integrated
  sensing and communications,'' \emph{arXiv preprint arXiv:2307.02242}, 2023.

\bibitem{Shao2022TargetSensing}
X.~Shao, C.~You, W.~Ma, X.~Chen, and R.~Zhang, ``Target sensing with
  intelligent reflecting surface: Architecture and performance,'' \emph{{IEEE}
  J. Sel. Areas Commun.}, vol.~40, no.~7, pp. 2070--2084, Jul. 2022.

\bibitem{Meng2022Intelligent}
K.~Meng, Q.~Wu, R.~Schober, and W.~Chen, ``Intelligent reflecting surface
  enabled multi-target sensing,'' \emph{{IEEE} Trans. Commun.}, vol.~70,
  no.~12, pp. 8313--8330, Dec. 2022.

\bibitem{Song2023Intelligent}
X.~Song, J.~Xu, F.~Liu, T.~X. Han, and Y.~C. Eldar, ``Intelligent reflecting
  surface enabled sensing: Cramér-rao bound optimization,'' \emph{IEEE Trans.
  Signal Process.}, vol.~71, pp. 2011--2026, 2023.

\bibitem{Alegra2022Orthogonalization}
J.~V. Alegría and F.~Rusek, ``Channel orthogonalization with reconfigurable
  surfaces,'' in \emph{IEEE Proc. Globecom Workshops}, 2022, pp. 37--42.

\bibitem{Khan2023IntegrationNOMA}
W.~U. Khan, E.~Lagunas, A.~Mahmood, Z.~Ali, M.~Asif, S.~Chatzinotas, and
  B.~Ottersten, ``Integration of {NOMA} with reflecting intelligent surfaces: A
  multi-cell optimization with {SIC} decoding errors,'' \emph{IEEE Trans. Green
  Commun. Netw.}, vol.~7, no.~3, pp. 1554--1565, 2023.

\bibitem{Foundations2022Buzzi}
S.~Buzzi, E.~Grossi, M.~Lops, and L.~Venturino, ``Foundations of {MIMO} radar
  detection aided by reconfigurable intelligent surfaces,'' \emph{IEEE Trans.
  Signal Process.}, vol.~70, pp. 1749--1763, 2022.

\bibitem{Shao2023National}
X.~Shao and R.~Zhang, ``{Enhancing wireless sensing via a target-mounted
  intelligent reflecting surface},'' \emph{Natl. Sci. Rev.}, vol.~10, no.~8,
  Jul. 2023.

\bibitem{Liu2022Integrated}
F.~Liu, Y.~Cui, C.~Masouros, J.~Xu, T.~X. Han, Y.~C. Eldar, and S.~Buzzi,
  ``Integrated sensing and communications: Toward dual-functional wireless
  networks for {6G} and beyond,'' \emph{{IEEE} J. Sel. Areas Commun.}, vol.~40,
  no.~6, pp. 1728--1767, Jun. 2022.

\bibitem{Fang2021OvertheAir}
W.~Fang, Y.~Jiang, Y.~Shi, Y.~Zhou, W.~Chen, and K.~B. Letaief, ``Over-the-air
  computation via reconfigurable intelligent surface,'' \emph{{IEEE} Trans.
  Commun.}, vol.~69, no.~12, pp. 8612--8626, Dec. 2021.

\end{thebibliography}
\bibliographystyle{IEEEtran}

\end{document}